\documentclass{article}

\def\giorno{29 October 2009}

\def\b{\beta}

\def\ga{\gamma}
\def\eps{\varepsilon}
\def\la{\lambda}
\def\s{\sigma}

\def\z{\zeta}

\def\vth{\vartheta}

\def\De{\Delta}
\def\Ga{\Gamma}
\def\La{\Lambda}
\def\phi{\varphi}

\def\G{{\mathcal G}}
\def\E{{\mathcal E}}

\def\L{{\mathcal L}}
\def\T{{\rm T}}
\def\G{{\mathcal G}}

\def\pa{\partial}

\def\d{{\rm d}}       

\def\ss{\subset}
\def\sse{\subseteq}

\def\({\left(}
\def\){\right)}
\def\[{\left[}
\def\]{\right]}
\def\~#1{\widetilde #1}
\def\.#1{\dot #1}
\def\^#1{\widehat #1}

\def\interno{\hskip 2pt \vbox{\hbox{\vbox to .18
truecm{\vfill\hbox to .25 truecm
{\hfill\hfill}\vfill}\vrule}\hrule}\hskip 2 pt}

\def\mapright#1{\smash{\mathop{\longrightarrow}\limits^{#1}}}
\def\mapdown#1{\Big\downarrow\rlap{$\vcenter{\hbox{$\scriptstyle#1$}}$}}
\def\mapleft#1{\smash{\mathop{\longleftarrow}\limits^{#1}}}

\def\mapse#1{\smash{\mathop{\searrow}\limits^{#1}}}

\def\beq{\begin{equation}}
\def\eeq{\end{equation}}

\def\qd{{\dot q}}
\def\xd{{\dot x}}
\def\yd{{\dot y}}
\def\zd{{\dot z}}

\def\xdd{{\ddot x}}
\def\ydd{{\ddot y}}
\def\zdd{{\ddot z}}

\def\={\, =\, }

\def\eqref#1{(\ref{#1})}

\begin{document}

\title{Twisted symmetries and integrable systems}

\author{Giuseppe Gaeta\\ {\it Dipartimento di Matematica} \\
{\it Universit\`a degli Studi di Milano,} \\ {\it via Saldini 50,
20133 Milano (Italy)} \\ {\tt giuseppe.gaeta@unimi.it}
\\ {} \\
Giampaolo Cicogna\\ {\it Dipartimento di Fisica, Universit\`a di
Pisa}  \\ and {\it INFN, Sezione di Pisa,} \\
{\it Largo B. Pontecorvo 3, 50127 Pisa (Italy)} \\ {\tt
cicogna@df.unipi.it} }

\date{\giorno}

\maketitle

\noindent {\bf Summary.} Symmetry properties are at the basis of
integrability. In recent years, it appeared that so called {\it
twisted symmetries} are as effective as standard symmetries in
many respects (integrating ODEs, finding special solutions to
PDEs). Here we discuss how twisted symmetries can be used to
detect integrability of Lagrangian systems which are not
integrable via standard symmetries.

\medskip
\noindent{\tt Keywords}: Symmetry of differential equations;
integrable systems; conservation laws.

\section*{Introduction}

Integrable systems are characterized by a high degree of symmetry
-- and a favourable structure of the underlying symmetry algebra.

In recent years, the standard concept of symmetry for differential
equations \cite{EMS,ArnGMDE,CGspri,Gbook,KrV,Olv1,Olv2,Ste,Win}
has been generalized in several directions (see e.g. \cite{CRC}
for an overview). Here we are interested in a special case among
these generalizations; actually this is special not only in the
sense it is a specific one but also in that it differs from all
other ones in a substantial way. That is, in dealing with symmetry
of differential equations we always consider a vector field acting
in the basic (independent and dependent) variables, and then {\it
prolong}  it to derivatives of suitable order -- the order of the
differential equation to be considered, or maybe to infinite
order. While ``usual'' generalizations amount to generalize the
admitted vector fields acting on basic variables, {\bf twisted
symmetries} modify the prolongation operation itself. In this note
we will focus specifically on twisted symmetries, and their role
in analyzing integrability.

Our main result will be that systems which are characterized by a
high degree of twisted symmetry -- and a favorable structure of
the underlying symmetry algebra -- are integrable.

Twisted symmetries are not a new concept, but a collective name to
include different types of symmetries ($C^\infty$-symmetries, also
known as $\lambda$-symmetries, $\mu$-symmetries, and the recently
introduced $\rho$-symmetries), all of them involving a deformation
of the prolongation operation. They were first introduced in the
context of scalar ODEs as $C^\infty$-symmetries or
$\lambda$-symmetries in a seminal paper by Muriel and Romero
\cite{MuRom1}, who also extended their ideas to more general
settings \cite{MuRomSPT,MuRom2,MuRom2b,MuRomVigo,MuRomJLT,
MuRomTMP,MuRom2007,MuRomIF,MuRom2009,MRO}; see also
\cite{GMM,PuS}. The name ``$\la$-symmetries'' refers to the key
role played by a $C^\infty$ function $\la (x,u,u_x)$. The case
dealing with PDEs was then christened under the name of
$\mu$-symmetries \cite{CGM,GM} both for alphabetic continuity and
because in this context the key role is played by a semi-basic
matrix-valued one-form $\mu = \La_i \d x^i$. Still a different
name in use is that of ``$\rho$-symmetries'' for a specific class
of $\mu$-symmetries which allow reduction of system of ODEs
\cite{Cicrho,Cic09}. The basic ideas behind twisted symmetries as
well as these different special types of twisted symmetries will
be briefly reviewed in section \ref{sec:twistprol} below; a more
substantial review is provided in \cite{Gae09}.

\section{Notation}
\label{sec:notation}

We will start by fixing some general (standard) notation, to be
freely used in the following.

We will consider problems defined on a phase bundle $(M,\pi,B)$
with fiber $\pi^{-1} (x) = U$; here $B$ and $U$ are smooth real
manifolds of dimensions $p$ and $q$ respectively, and we will use
local coordinates $\{ x^1,...,x^p\}$ in $B$ and $\{ u^1,...,u^q\}$
in $U$. As usual, when dealing with differential equations we will
think of the $x$ as independent variables and the $u$ as dependent
ones (fields). Associated to the bundle $M$ are the $k$-th order
{\it jet bundles} $J^k M$; there are natural coordinates in these,
provided by $x,u$ and by partial derivatives of the $u$ with
respect to the $x$. In dealing with these, we will freely use the
multi-index notation, see e.g. \cite{Olv1} for
details.\footnote{In this note we will actually mainly focus on
systems with only one independent variable. However we prefer to
deal with the general case as this makes the geometry behind
twisted symmetries -- and their properties -- more transparent:
the special case of ODEs is indeed degenerate in several respects,
and it is highly remarkable that Muriel and Romero were able to
deal with it at first.}

\subsection*{Vector fields}

Consider now a vector field (VF) on $M$; this will be written in
coordinates as\footnote{Here and everywhere below we understand
summation over repeated indices unless otherwise stated.}
\beq\label{eq:vf} X \ = \ \xi^i (x,u) \, \frac{\pa}{\pa x^i} \ + \
\phi^a (x,u) \, \frac{\pa}{\pa u^a} \ . \eeq We will routinely
omit to write the dependencies of the functions (such as $\xi$ and
$\phi$) to avoid an exceedingly heavy notation.

As well known, in many cases it is convenient to consider the {\it
evolutionary representative} (or vertical representative) of a VF;
this describes the action of the VF on a section of the bundle
$(M,\pi,B)$ and is written in coordinates as\footnote{In fact, a
simple computation shows that if a section $\s$ is given in
coordinates by $\s = \{(x , u ) \ / \ u = f(x) \}$, then under the
infinitesimal action of $X$ it is mapped to a new section $\^\s =
\{(x , u ) \ / \ u = \^f (x) \}$ with $ \^f (x) = f(x) + \eps \[
\phi^a - \xi^i u^a_i \]_\s$, where the functions within the square
bracket should be computed on the section $\s$.}
\beq\label{eq:vvf} X_v \ = \ \( \phi^a \, - \, \xi^i \, u^a_i \) \
\frac{\pa}{\pa u^a} \ = \ Q^a \ \frac{\pa}{\pa u^a} \ . \eeq

A vector field acting in $M$ also acts naturally in $J^k M$: once
the action on independent and dependent variables is given, the
action on derivatives of any order can be readily computed. The
lift of the $X$ action from $M$ to $J^k M$ is also known as the
{\it prolongation} operation \cite{EMS,CGspri,Gbook,KrV,Olv1,
Olv2,Ste,Win}. In coordinates, the prolonged vector field $X^*$
(or $X^n$ if we consider prolongation only up to order $n$) is
given by \beq\label{eq:vfprol} X^* \ = \ \xi^i \, \frac{\pa}{\pa
x^i} \ + \ \psi^a_J  \, \frac{\pa}{\pa u^a_J} \ , \eeq where $J$
are multi-indices and $$ \psi^a_0 \ = \ \phi^a \ . $$ The
coefficients $\psi^a_J$ are then determined by the prolongation
formula \beq\label{eq:prolform} \psi^a_{J,i} \ = \ D_i \psi^a_J \
- \ u^a_{J,k} \, D_i \xi^k \ . \eeq Here $D_i$ is the total
derivative with respect to $x^i$, i.e. $$ D_i \ = \ \frac{\pa}{\pa
x^i} \ + \ u^a_{J,i} \, \frac{\pa}{\pa u^a_J} \ .
$$ These relations are specially simple for vertical vector
fields: in this case we have \beq\label{eq:vvfprol} X_v^* \ = \
Q^a_J \, \frac{\pa}{\pa u^a_J} \ , \ \ \mathrm{with} \ \ Q^a_{J,i}
\ = \ D_i Q^a_J \ . \eeq

It will be convenient to have a more intrinsic characterization of
the prolongation operation.

The jet bundles $J^n M$ are naturally equipped with a {\bf contact
structure}, i.e. with a set of contact forms $\theta^a_J$, given
in coordinates by \beq\label{eq:cont} \theta^a_J \ := \ \d u^a_J \
- \ u^a_{J,i} \, \d x^i \ \ \ \ (|J| = 0,...,n-1) \ . \eeq We will
denote by $\E$ the ideal generated by these forms (with
coefficients in $C^\infty (J^n M)$).

Then the prolonged vector field $X^* = Y$ (we use this notation
for graphical ease) is the only vector field which coincides with
$X$ on $M$ and which {\it preserves} the contact ideal $\E$, i.e.
such that \beq\label{eq:contgeom} L_{Y} (\E) \ \in \ \E \ ; \eeq
here $L$ is the Lie derivative. This means that for any $\vartheta
\in \E$, $L_{Y} (\vartheta) \in \E$.

The condition \eqref{eq:contgeom} can be expressed equivalently in
terms of conditions involving the commutator of $Y$ with the total
derivative operators $D_i$; in particular, it is equivalent to
either one of \begin{eqnarray}
 & & [D_i , Y ] \interno \vth \ = \ 0 \ \ \ \forall \vth \in \E \
 , \label{eq:ceq1} \\
 & & [D_i , Y ] \ = \ h_i^m \, D_m \ + V \ , \label{eq:ceq2}
\end{eqnarray} with $h_i^m \in C^\infty (J^n M)$ and $V$ a vertical vector field
in $J^n M$ seen as a bundle over $J^{n-1} M$ (i.e. it has
components only along derivatives of maximal order $n$).

It is appropriate, for further reference and since it has just
been mentioned, to recall that the jet bundles have several
fibered structures; in particular, $J^k M$ can be seen both as a
bundle $(J^k M, \pi_k , B)$ over $B$ and as a bundle
$(J^k,\s_k,M)$ over $M$: $$ \matrix{J^k M & & \cr & & \cr
\mapdown{\pi_k} & \mapse{\s_k} & \cr & & \cr B & \mapleft{\pi} & M
  \cr}  $$

\subsection*{Differential equations and their symmetries}

A differential equation -- or system of differential equations --
of order $n$, which we will denote by $\Delta$, identifies a
submanifold $S_\Delta \ss J^n M$, its solution manifold. That is,
$S_\De$ is the set of points of $J^n M$ in which the relations
$\Delta$ are satisfied. If the equations involve only smooth
coefficients, then $S_\De$ is smooth, and we will assume that the
equation is non-degenerate\footnote{That is, if $\Delta$ is given
by $E^a (x,u^{(n)} ) = 0$, we assume that the derivatives of $E^a$
are nonzero in directions transversal to $S_\De$}, so that they
correspondence between $\De$ and $S_\De$ is one-to-one
\cite{Olv1}.

The vector field $X$ is a symmetry (or more precisely, being a VF,
a symmetry generator) for $\De$ if its $n$-th prolongation
$X^{(n)}$ leaves $S_\De$ invariant, i.e.
$$ X^{(n)} \ : \ S_\De \, \to \, \T \, S_\De \ . $$
In this case, the one-parameter group generated by $X$ maps
solutions to $\De$ into solutions.

We assume the reader is familiar with the use of symmetries for
the analysis of differential equations, referring to
\cite{EMS,Gbook,KrV,Olv1,Olv2,Ste} for details and applications;
here we just wanted to stress that the very concept of symmetry of
differential equations is based on the prolonged vector fields and
hence on the prolongation operation.

\section{Twisted prolongations and symmetries}
\label{sec:twistprol}

We will now introduce and discuss the twisted prolongation $Y$ of
a vector field $X$; we anticipate that if $Y$ satisfies the
symmetry condition \beq Y \ : \ S_\De \, \to \, \T \, S_\De \ ,
\eeq then $X$ is a twisted symmetry of the differential equation
$\De$.

Here we will deal directly with the most general setting ($p$
independent and $q$ dependent variables for a system of PDEs of
order $n$); we refer to \cite{Gae09} for a review and discussion
of relevant special cases: e.g. scalar equations, or ODEs; the
latter will also be discussed below.

The relevant contact structure in this case is spanned by the
contact forms \eqref{eq:cont}; it is convenient to see them as the
components of a vector-valued contact form $\vth_J$ \cite{Str}. We
will denote by $\Theta$ the module over $q$-dimensional smooth
matrix functions generated by the $\vth_J$, i.e. the set of
vector-valued forms which can be written as $\eta = (R_J)^a_b
\vth^b_J $ with $R_J : J^n M \to \mathrm{Mat}(q)$ smooth matrix
functions.

The manifold of dependent variables -- that is, the fiber of
$(M,\pi,B)$ -- has tangent space $U \simeq R^q$, on which is
defined an action of $G = GL(q,R)$; the corresponding Lie algebra
is $\G = g \ell (q) $ (we omit from now on the indication that all
our manifolds, spaces and actions are real).\footnote{We will
think the $G$ and hence $\G$ action is fixed once for all, and
hence -- for the sake of notation -- do not distinguish
notationally between the group (or algebra) and its
representation.}

Consider a $\G$-valued semi-basic one-form on $J^1 M$,
\beq\label{eq:mu} \mu \ := \ \La_i \, \d x^i \ ; \eeq the $\La_i =
\La_i (x,u,u_x)$ are smooth matrix functions (with values in $\G$)
satisfying some additional compatibility conditions discussed
below, see \eqref{eq:comp}.

We will say that the vector field $Y$ on $J^n M$ {\it
$\mu$-preserves } the vector contact structure $\Theta$ if, for
all $\vth \in \Theta$, \beq \L_Y (\vth ) \ + \ \( Y \interno
(\La_i)^a_b \vth^b \) \ \d x^i  \ \in \ \Theta \ ; \eeq this
should be compared to standard preservation of the contract
structure in the form \eqref{eq:contgeom}.

In terms of the coefficients of $Y$, see \eqref{eq:vfprol}, this
is equivalent to the requirement that the $\Psi^a_J$ obey the {\bf
vector $\mu$-prolongation formula} \beq\label{eq:muprol}
\Psi^a_{J,i} \ = \ (\nabla_i)^a_b \, \Psi^b_J \ - \ u^b_{J,m} \,
[( \nabla_i)^a_b \, \xi^m ] \ , \eeq where we have introduced the
(matrix) differential operators
$$ \nabla_i \ := \ I \, D_i \ + \ \La_i \ ;  $$
here $I$ is of course the $q \times q$ identity matrix. Needless
to say, for $\mu = 0$ (i.e. $\La_i = 0$ for all $i$), this reduces
to the standard prolongation formula.

Note for later reference that in the case of vertical vector
fields $X = Q^a (\pa / \pa u^a)$, \eqref{eq:muprol} yields for the
coefficients of the first prolongation $Y = X + \psi^a_i (\pa /
\pa u^a_i)$, simply  \beq\label{eq:vproluno} \psi^a_i \ = \
(\nabla_i)^a_b \, Q^b \ = \ D_i \, Q^a \ + \ (R_i)^a_{\ b} \, Q^b
\ . \eeq

As mentioned above, the functions $\La_i$ defining the form $\mu$
in \eqref{eq:mu} are not arbitrary: they must satisfy a
compatibility condition (this guarantees the $\Psi^a_J$ defined by
\eqref{eq:muprol} are uniquely determined), expressed in
coordinates by \beq\label{eq:comp}
\[ \nabla_i , \nabla_k \] \ \equiv \ D_i \La_k \, - \, D_k \La_i \ + \
[\La_i , \La_k ] \ = \ 0 \ . \eeq

This is nothing else than the coordinate expression of the
horizontal Maurer-Cartan equation\footnote{This expresses the
requirement that the standard Maurer-Cartan equation is satisfied
modulo contact forms, i.e. $\d \mu + (1/2) [\mu , \mu] \in \E$.}
\cite{CGM} \beq\label{eq:hMC} D \mu \ + \ {1 \over 2} \ [ \mu ,
\mu ] \ = \ 0 \ . \eeq Based on this condition -- and on classical
results of differential geometry \cite{EGH,CCL,Sha} -- it follows
easily that in any contractible neighborhood $A \sse J^n M$, there
exists $\ga_A : A \to GL(q)$ such that (locally in $A$) $\mu$ is
the Darboux derivative of $\ga_A$.

In other words, any $\mu$-prolonged vector field is {\it locally}
gauge-equivalent to a standard prolonged vector field
\cite{CGM,Mor2007}, the gauge group being $G=GL(q)$.\footnote{When
$J^n M$ is topologically nontrivial, or $\mu$ presents singular
points, one can have $\mu$-prolonged vector fields which are not
globally gauge equivalent to standardly prolonged ones (and in
this sense non-trivial $\mu$-symmetries); see \cite{CGM} for
concrete examples.}

The result stated above means that if $Y$ is the
$\mu$-prolongation of a vector field $X$, then there are vector
fields $W$ and $Z$, gauge-equivalent via the same gauge
transformation (acting respectively as $\ga^{(k)}$ in $\T (J^k) M$
and as $\ga$ in $\T (M)$) to $Y$ and $X$, and such that $W$ is the
standard prolongation of $Z$. This is schematically summarized in
the following diagram:
$$ \matrix{ X & & \mapright{\ga} & & Z \cr
 & & & & \cr
\mapdown{\mu-{\mathrm{prol}}} & & & & \mapdown{{\mathrm{prol}}}
\cr & & & & \cr Y & & \mapright{\ga^{(k)}} & & W \cr}  $$ For
these considerations, it is convenient to deal with evolutionary
representatives of vector fields \cite{Olv1}, which we will
implicitly do from now on.

It should also be stressed that the gauge group $\Ga$ (modelled
over a Lie group $G$) acts in the same way on the vector $\{
\phi^1 , ... , \phi^q \}$ of the components of the vector field
$X$ in $M$, and on the vectors $\{ \psi^1_J , ... , \psi^q_J \}$
of components (relative to a given multi-index $J$, i.e. to
partial derivatives with respect to the same array of independent
variables) of the vector field $Y$ in $J^k M$. One also says that
$\Ga$ acts via a {\it Jet representation}.

Summarizing, one finds out that the twisted prolongation operation
amounts locally to standard prolongation seen in a different
reference frame, i.e. under a gauge transformation.\footnote{This
point of view was first presented in \cite{CGM,GM} and is further
discussed in \cite{Ggau1,Ggau2,Gae09}.}

It appears now fully naturally that -- for what concerns
properties which are both {\it local } and {\it frame-independent
} -- twisted symmetries are as good as standard ones.

The existence of conserved quantities in Mechanics -- or conserved
currents in Field Theory -- satisfies these criteria, and it
should thus be no surprise that one can analyze these with the
help of twisted symmetries as well as of standard ones.

\section{Lagrangians, twisted symmetries and conservation laws}
\label{sec:lagrangians}

The reader who attempts some very simple computations could find
very strange the above statement, that twisted symmetries of the
Lagrangian are related to conserved quantities. In fact, it is
easy to check -- e.g. in the case of first order Lagrangians $\L
(q, \qd;t)$ which we will consider for the sake of simplicity --
that in general a twisted symmetry of the Lagrangian $\L$ does not
correspond to a symmetry, either regular or twisted, of the
corresponding Euler-Lagrange equations \beq\label{eq:EL} \frac{\pa
L }{\pa q^i} \ - \ \frac{d}{d t} \, \frac{\pa \L}{\pa \qd^i} \ = \
0 \ ; \eeq and even less to a conserved
quantity.\footnote{Needless to say, the same is true {\it a
fortiori } in the case of a field Lagrangian $\L (u , u_x ;
x^1,..., x^d)$ and the corresponding Euler-Lagrange equations $
(\pa \L / \pa u^a) - (d / d x^i) (\pa \L / \pa u^a_i) = 0$: no
conserved quantity is in general associated to a twisted symmetry
of the Lagrangian. See \cite{CGnoe} for a detailed discussion.}

The point is that if we change reference frame acting on the
variables $q^i$ via a gauge transformation $\La^i_{\ j}$, the
variational equations corresponding to the Lagrangian $\L (q,
\qd;t)$ are not the standard Euler-Lagrange equations
\eqref{eq:EL}, but rather the ``twisted Euler-Lagrange equations''
\beq\label{eq:ELT} \frac{\pa L }{\pa q^i} \ - \ \frac{d}{d t} \,
\frac{\pa \L}{\pa \qd^i} \ + \ \frac{\pa \L}{\pa \qd^j} \,
\La^j_{\ i} \ = \ 0 \ ; \eeq for a derivation and discussion of
these equations, see \cite{CGnoe}.

\medskip\noindent
{\bf Remark.} Albeit we decided to restrict in general to the
Mechanics case, we would now like to mention that in the general
field theoretical case, with fields $u^a$ and space-time variables
$x^i$, we have several matrix functions $\La_i$ (one for each
space-time variable), subject to the compatibility conditions
discussed in section \ref{sec:twistprol} (which we assume are
satisfied); introducing the notations $\pi_a^i := (\pa \L / \pa
u^a_i )$ for the momenta and $D_i$ for the total derivative with
respect to the variable $x^i$, the resulting equations read \beq
\label{eq:ELTg} \frac{\pa L }{\pa u^a} \ - \ D_i \, \pi_a^i \ = \
- \, \( \La^T_i \)_a^{\ b} \, \pi_b^i \ . \eeq The reader is again
referred to \cite{CGnoe} for a derivation. \hfill $\odot$
\bigskip

It is then a simple matter to check that the following statement
(which is Theorem 9 from \cite{CGnoe}) holds true:

\medskip\noindent
{\bf Proposition 1.} {\it Let $\L$ be a first order field
Lagrangian, admitting the vector field $X = \varphi^a (\pa / \pa
u^a)$ as a $\mu$-symmetry for a certain form $\mu = \La_i \d x^i$.
Then the vector ${\bf P}$ of components ${\bf P}^i = \varphi^a
\pi^i_a$ defines a standard conservation law, $D_i {\bf P}^i = 0$,
for the flow of the associated $\mu$-Euler-Lagrange equations
\eqref{eq:ELTg}.}
\bigskip

In the case of a Mechanical Lagrangian $\L (q,\qd;t)$, the above
Proposition 1 reduces to a statement about the existence of first
integrals:

\medskip\noindent
{\bf Proposition 2.} {\it Let $\L (q,\qd;t)$ be a first order
mechanical Lagrangian, admitting the evolutionary vector field $X
= \varphi^i (\pa / \pa q^i)$ as a $\mu$-symmetry for a certain
form $\mu = \La \d t$. Then the function $J = \varphi^i ( \pa \L /
\pa \qd^i)$ is a first integral for the flow of the associated
$\mu$-Euler-Lagrange equations \eqref{eq:ELT}, i.e. $d J / d t =
0$.}
\bigskip

The reader is once again referred to \cite{CGnoe} for applications
and examples\footnote{And also for a discussion of how twisted
symmetries induce ``$\mu$-conservation laws'' for the standard
Euler-Lagrange equations, and correspondingly how standard
symmetries induce the same kind of ``$\mu$-conservation laws'' for
the twisted Euler-Lagrange equations.}. See also the earlier paper
\cite{MRO}, which started application of twisted symmetries to the
study of variational problems.

\section{Multiple twisted symmetries and reduction}
\label{sec:lie}

As we have discussed in the previous Section
\ref{sec:lagrangians}, each twisted symmetry of the Lagrangian
yields a first integral and hence allow for a reduction of the
variational problem. If we have several symmetries, we can try to
use them one after the other to reduce by stages the variational
problem \cite{CMOR}; however, this will be effective only if the
Lie algebra of vector fields generating the Lagrangian symmetries
have a convenient structure. This corresponds to what happens in
the usual application of Lie symmetries to differential equations,
where in general only a solvable algebra can be fully used for
reduction.\footnote{It should be mentioned that after reduction
one could have extra symmetries beyond those inherited from the
unreduced problem, see e.g. the discussion in \cite{Olv1}; this
remark \cite{Bas} was one of the motivations for the introduction
of twisted symmetries, and on the other hand leads to considering
{\it solvable structures} \cite{BPr,SPr}. See \cite{CFM} for a
recent discussion blending (twisted) symmetries and solvable
structures in the reduction of ODEs.}

Needless to say, the same will hold in this case. However, the
situation here is slightly more complex, and could appear much
more complex if one is not aware of the basic mechanism at work,
i.e. that twisted symmetries correspond -- locally -- to standard
symmetries in a different reference frame. Note that as the Lie
algebraic structure of sets of vector fields depends only on local
properties, we can make full use of this feature in the present
context.

We stress that the twisting should be {\it the same for all vector
fields}, i.e. we operate with the same matrix $\La$ for the
prolongation of different vector fields; this correspond to the
fact that the associated gauge transformation is the same.

Let us denote the vector fields generating twisted symmetries as
\beq\label{eq:XVF11} X_a \ = \ \varphi^i_a \ (\pa / \pa q^i ) \ ,
\eeq and their first prolongation as $Y_a$. We have \beq Y_a \ = \
X_a \ + \ \psi^i_a \ (\pa / \pa \qd^i ) \ ; \eeq we recall that
\beq\label{eq:prol11} \psi^i_a \ = \ D_t \varphi^i_a \ + \
\La^i_{\ j} \, \varphi^j_a \ . \eeq

Let us determine the reference frame in which prolongations are
just standard ones. Acting on component of vector fields by an
invertible matrix function\footnote{We recall that in general $R$
can depend not only on the independent variable $t$, but in the
dependent variables $q^i$ as well.} $R$, and writing $\varphi = R
\xi$, $\psi = R \eta$, eq.\eqref{eq:prol11} reads
$$ R \, \eta \ = \ D_t \, (R \, \xi) \ + \ \La \, R \, \xi \ , $$
which also reads
$$ \eta \ = \ D_t \, \xi \ + \ \[ (R^{-1} \, D_t R) \ + \ (R^{-1} \La  R) \]
\, \xi \ . $$ Thus the components of the prolonged vector field
satisfy the standard prolongation formula (for vertical vector
fields) $\eta = D_t \xi$ if and only if $R$ and $\La$ are related
by \beq\label{eq:gaugetr} (R^{-1} \, D_t R) \ + \ (R^{-1} \La  R)
\ = \ 0 \ . \eeq Needless to say, this just expresses the request
that the gauge transformation $R$ maps $\La$ into the identically
null matrix function, as the left hand side of \eqref{eq:gaugetr}
is the standard expression for a gauge transformation, see e.g.
\cite{CCL,EGH,Nak,NaS}. We thus have \beq R \ = \ \exp \[ - \int
\La \, \d t \] \ . \eeq

The commutator of vector fields $X_a$, $X_b$ as above is given by
$$ \[ X_a , X_b \] \ = \ \{ \varphi_a , \varphi_b \}^i \ (\pa /
\pa q^i ) \ , $$ where we have defined \beq\label{eq:LPP} \{
\varphi_a , \varphi_b \}^i \ := \ \varphi^j_a \, \frac{\pa
\varphi^i_b}{\pa q^j} \ - \ \varphi^j_b \, \frac{\pa
\varphi^i_a}{\pa q^j} \ . \eeq

When we see this as twisted vector fields, i.e. vector fields on
which the gauge transformation $R$ acted, we rewrite $\varphi = R
\xi $ and hence\footnote{By performing the corresponding
transformation $q = R \chi$ on the dependent variables we would
get rid of $R$, but here we want to briefly discuss how the
algebraic relations between the $X_a$ are affected by this
transformation. This is easily done by using the bracket $\{ . , .
\}$ defined above, and a generalization to be defined in a moment.
} $$ X_a \ = \ (R^i_{\ j} \xi^j ) \ (\pa / \pa q^i) \ . $$

In this way we get immediately
$$ \[ X_a , X_b \] \ = \ \{ R \xi_a , R \xi_b \}^i \ (\pa /
\pa q^i ) \ := \ \{ \xi_a , \xi_b \}^i_{(R)} \ (\pa / \pa q^i ) \
. $$ We can summarize this relation, with the notations introduced
above, as $$ \{ \varphi_a , \varphi_b \} \ = \ \{ \xi_a , \xi_b
\}_{(R)} \ ; $$ this also reads $$ \{ \varphi_a , \varphi_b
\}_{(R^{-1})} \ = \ \{ \xi_a , \xi_b \} \ . $$ Now we note that
the $\xi$ correspond to the reference frame in which prolongations
are not twisted, i.e. $\La = 0$, and the Euler-Lagrange equations
are the standard ones. In this frame a solvable Lie algebra of
vector fields allows for reduction by stages, and the structure of
the Lie algebra is recovered by considering $\{ \xi_a , \xi_b \}$.

Thus, if the twisted symmetries $X_a$ of $\L$, written in the form
\eqref{eq:XVF11}, are obtained by twisting the prolongation by
$\La$, one determines the corresponding $R$ and should then check
that the vector of vector fields component $\varphi$ form a
solvable Lie algebra with the bracket \beq\label{eq:RLB} \{ . , .
\}_{(R^{-1})} \ . \eeq

If the maximal solvable algebra (under this bracket) of twisted
symmetries has dimension $n$, then the system described by the
$n$-dimensional Lagrangian $\L$ is integrable.

Finally, we would like to stress that we conducted our discussion
locally; Integrability is however usually of interest when is a
global property, so that one should be able to patch together the
analysis in different local charts to extend it over the whole
manifold. In this sense the analysis in terms of the bracket
\eqref{eq:RLB} is more convenient -- at least notationally -- than
simply passing to the gauge transformed frame (in which the
twisted prolongation is mapped to a regular one), in that
different gauge transformations and hence different $R$ would be
used in different local charts.


\section{Discussion and outlook}

In this final section we collect some remarks on the subject
discussed here as well as comments about devisable further
developments.

\medskip\noindent
{\bf (1)} The point of view adopted here is to use as far as
possible the result that {\it locally } twisted symmetries are
standard ones described in a different (twisted) frame of
reference, i.e. deformed by a gauge transformation. In this way,
many of the results obtained for twisted symmetries appear more
and less obvious, and focus should be shifted to global
properties. Needless to say, this also applies to Integrability;
we have focused on the local aspects in that the transition from
local to global ones does not present any special feature in the
case of twisted symmetries, except in some respect for what
concerns the analysis of the Lie algebraic properties of sets of
different twisted symmetries; for these we have suggested in
Section \ref{sec:lie} a way to analyze the situation in terms of a
deformed bracket which takes into account the twisting in
different local charts, i.e. under the different local gauge
transformations mapping twisted prolongations into standard ones.

\medskip\noindent
{\bf (2)} The relation between twisted symmetries and
gauge-transformed standard ones also suggests that one could
proceed in a different way. That is, the problem here arises from
the fact that in considering standard (partial or ordinary)
derivatives, as common in Applied Mathematics, they transform in a
different way when one changes the reference frame. The cure for
this is well known in Physics, and consists in using {\it
covariant derivatives}. One could thus consider equations in terms
of covariant derivatives, and prolong vector fields acting in the
phase manifold $M$ by considering their action on covariant
(rather than standard) derivatives, i.e. by employing ``covariant
jet spaces''. We hope this approach can be implemented in future
works.

\medskip\noindent
{\bf (3)} Working instead with standard derivatives, it is quite
clear that the approach based on gauge transformations allows to
easily produce integrable systems simply by starting from a known
one and applying gauge transformations on it. The transformed
systems will in general not be integrable in usual sense (that is,
not pass the usual integrability tests), in particular not possess
a suitable algebra of standard symmetries, but will -- just by
construction ! -- possess a suitable algebra of twisted symmetries
and be integrable in the sense considered in the present paper. It
goes without saying that albeit here we considered Lagrangian
systems only, this remark applies to any kind of Integrable
System.

\medskip\noindent
{\bf (4)} It is also rather clear that albeit we only discussed
first order mechanical Lagrangians, the geometrical framework and
hence the validity of the present approach are quite more general.
In particular they also apply both to higher order Lagrangians and
to the framework of Field Theory.

\medskip\noindent
{\bf (5)} Finally we recall that, as already mentioned above, the
Hamiltonian aspects of this approach have received only limited
attention, and only some preliminary results are at present
available.

\bigskip\noindent

These remarks suggest directions for further developments. We hope
some of the readers the present paper will contribute to these.

\section*{Appendix. A concrete example}

In order to illustrate our discussion, we discuss a fully explicit
(but slightly artificial) example, considering a mechanical
Lagrangian in three degrees of freedom; we denote by $t$ the
independent variable and by $(x,y,z)$ the dependent ones.

\subsection*{A.1 The Lagrangian and its twisted symmetries}

We choose the Lagrangian $$ \label{exa:lag} \L \ = \ (1/2) \ [\zd
+ (x^2 + y^2) \, z \, f\((x^2+y^2) z\) ] \ \( \xd^2 + \yd^2 \) \ -
\ g \( (x^2+y^2) z \) \ . \eqno(A.1) $$ Here $f$ and $g$ are
arbitrary smooth functions of their argument $\b = [(x^2+y^2) z]$,
and we will assume $f \not= 0$ (in order to avoid discussing
trivial cases). The case $g=0$ is simple but not trivial, and the
reader willing to consider a specially simple example at first can
set $g=0$ in all the following formulas.

Let us now consider the two vector fields $$ \label{exa:vf} X \ =
\ x \pa_y \, - \, y  \pa_x \ , \ \ Y \ = \ x \pa_x \, + \, y \pa_y
\, - \, 2 z \pa_z \ . \eqno(A.2) $$ Any smooth function of $\b$ is
invariant under both of these, so in particular the potential part
$g(\b)$ of the Lagrangian $\L$, as well as the coefficient $[ \b
f(\b) ]$, are surely invariant. Note moreover that $X$ and $Y$
commute, $[X,Y]=0$.

As for the prolongations of $X$ and $Y$, we start by considering
their standard prolongations, which are easily seen to be
$$ \label{exa:standprol} X^{(1)} = X - \yd \pa_{\xd} + \xd \pa_{\yd}
\ , \ \ Y^{(1)} = Y + \xd \pa_{\xd} + \yd \pa_{\yd} - 2 \zd
\pa_{\zd} \ . $$ Applying these on the Lagrangian (A.1) we see
easily that $$ X^{(1)} \cdot \L \ = \ 0 \ , \ \ Y^{(1)} \cdot \L \
= \ [(x^2+y^2) (\xd^2+\yd^2) ] \, z \, f((x^2+y^2) z) \ \not= \ 0
$$ (the last inequality depends on the assumption $f \not= 0$). It
is easily checked that -- as also guaranteed by a general theorem
\cite{Olv1} given that the vector fields themselves commute --
these standard prolongations commute, $[X^{(1)},Y^{(1)} ] = 0 $.

Let us now consider twisted prolongation, with the twisting matrix
$$ \label{exa:Lambda} \La \ = \ (x^2+y^2) \ f((x^2+y^2) z) \
\pmatrix{ 0&0&0\cr 0&0&0\cr 0&0&1\cr} \ . $$ In this case we get,
applying the general formulas, $$ \label{exa:twistprol}
\begin{array}{l} X^{(1)}_\La \ = \ X \ - \yd \pa_\xd + \xd \pa_\yd \ , \\
Y^{(1)}_\La \ = \ Y \ + \xd \pa_\xd + \yd \pa_\yd - 2 (\zd +
(x^2+y^2) z f ((x^2+y^2) z) \pa_\zd \ . \end{array} $$ In this
case, the twisted prolongations do still commute,
$[X^{(1)}_\La,Y^{(1)}_\La ] = 0 $.

By applying these twisted prolongations to $\L$, we get $$
X^{(1)}_\La \cdot \L  \ = \ 0 \ , \ \ Y^{(1)}_\La \cdot \L \ = \ 0
\ ; $$ we thus have a two-dimensional (abelian) algebra of twisted
symmetries for the three-degrees-of-freedom Lagrangian $\L$, and
our results apply.

\subsection*{A.2 Conservation laws}

We want to check in particular that the conservation of quantities
$J_X$ and $J_Y$ (associated to the vector fields $X$ and $Y$
respectively) under the twisted Euler-Lagrange equations of
motion, granted by Proposition 2, holds in this case.

As easy to foresee in view of our choice for $\La$, the twisted
Euler-Lagrange equations \eqref{eq:ELT} are only slightly more
complex that the standard ones \eqref{eq:EL}, and actually differ
from these only for the equation related to $z$. It will be
convenient to introduce a simplified notation, with
$$ r^2 = x^2 + y^2  ,  \ \rho^2 = \xd^2 + \yd^2  ; \ \
F := f((x^2+y^2) z) , \ G = g ((x^2+y^2) z) \ . $$ In this way the
twisted Euler-Lagrange equations for $\L$ turn out to be $$
\label{exa:TEL} \begin{array}{l} -2 G_\b x z - 2 F \xd y \yd z - F
x (\xd^2 - \yd^2) z + \xdd (-F x^2 z - F y^2 z - \zd) -
  F x^2 \xd \zd \\
  \ \ \ - F \xd y^2 \zd - F_\b r^2 z (x (\xd^2 - \yd^2) z + x^2 \xd \zd + \xd y (2 \yd z +
y \zd)) - \xd  \zdd \ = \ 0 \, \\
\ \ \ - 2 G_\b y z + F \xd^2 y z - 2 F x \xd \yd z - F y \yd^2 z +
  \ydd (-F x^2 z - F y^2 z - \zd) - F x^2 \yd \zd \\ \ - F y^2 \yd \zd -
  F_\b r^2 z (-\xd^2 y z + 2 x \xd \yd z + \yd (y \yd z + x^2 \zd + y^2 \zd)) - \yd
  \zdd\ = \ 0 , \\
-\xd \xdd - \yd \ydd + (1/2) (-2 G_\b r^2 + F r^2 \rho^2 + r^2
\rho^2 (F + F_\b r^2 z)) \ = \ 0 \ . \end{array} $$

After some rearrangement, these provide $$ \label{exa:TELsol}
\begin{array}{rl} \xdd = & [\rho^2 (F r^2 z + \zd)]^{-1} \
\[(1/2) (2 F^2 r^4 \rho^2 \xd z + (-2 G_\b + F_\b r^2 \rho^2 z)
(2 x \yd^2 z \ \right. \\ &
 \left. \ + \xd (-2 y \yd z + x^2 \zd + y^2 \zd)) +
        F (F_\b r^6 \rho^2 \xd z^2 \right. \\ & \left. \ +
              2 (-G_\b r^4 \xd z +
                    \rho^2 (x \yd^2 z + \xd (-y \yd z + x^2 \zd + y^2
                    \zd)))))\] \ , \\
\ydd  = & - [\rho^2 (F r^2 z + \zd) ]^{-1} \ \[ -\xd (\xd y - x
\yd) z (-2 G_\b + F \rho^2 + F_\b r^2 \rho^2 z) \right. \\ &
\left. \ +
  r^2 (1/2) \yd (-2 G_\b + 2 F \rho^2 + F_\b r^2 \rho^2 z) (-F r^2 z - \zd) \] \
  , \\
\zdd  = & - [2 \rho^2 ]^{-1} \ \[ (2 F^2 r^4 \rho^2 z - 2 G_\b (-2
x \xd z + x^2 \zd + y (-2 \yd z + y \zd)) \right. \\ & \left. \ +
    F_\b r^2 \rho^2 z (2 x \xd z + 3 x^2 \zd + y (2 \yd z + 3 y \zd)) +
    F (F_\b r^6 \rho^2 z^2 \right. \\ & \left. \ +
          2 (-G_\b r^4 z + \rho^2 (x \xd z + 2 x^2 \zd + y (\yd z + 2 y \zd)))) \]
\ . \end{array} \eqno(A.3) $$

As for the conserved quantities associated to $X$ and $Y$
according to Proposition 2, these are respectively $$
\begin{array}{l} J_X \ = \ x \, \yd \, (F r^2 z + \zd) \ - \
y \, \xd \, (F r^2 z + \zd) \ , \\
J_Y \ = \  x \, \xd \, (F r^2 z + \zd) \ + \ y \, \yd \, (F r^2 z
+ \zd) \ - \ \rho^2 z \ . \end{array} $$

Their time derivatives are respectively $$
\begin{array}{rl} D_t (J_X) = & -2 F x \xd^2 y z + 2 F x^2 \xd \yd z - 2 F \xd y^2 \yd z + 2 F x y \yd^2 z -
  F x^2 \xd y \zd \\ & - F \xd y^3 \zd + F x^3 \yd \zd + F x y^2 \yd \zd +
  \ydd (F x^3 z + F x y^2 z + x \zd) \\ & + \xdd (-F x^2 y z - F y^3 z - y \zd) +
  F_\b r^2 (-\xd y + x \yd) z (2 x \xd z + x^2 \zd \\ &
  + y (2 \yd z + y \zd)) + (-\xd y +
        x \yd) \zdd           \ ; \\
  D_t (J_Y) = & 3 F x^2 \xd^2 z + F \xd^2 y^2 z + 4 F x \xd y \yd z + F x^2 \yd^2 z +
  3 F y^2 \yd^2 z \\
   & +
  F x^3 \xd \zd + F x \xd y^2 \zd + F x^2 y \yd \zd + F y^3 \yd \zd \\ & +
  \xdd (F x^3 z - 2 \xd z + F x y^2 z + x \zd) +
  \ydd (F x^2 y z + F y^3 z - 2 \yd z + y \zd) \\ & +
  F_\b r^2 (x \xd + y \yd) z (2 x \xd z + x^2 \zd + y (2 \yd z + y \zd)) + (x \xd +
        y \yd) \zdd \ . \end{array} $$

Inserting (A.3) into these, one checks that indeed $$ D_t (J_X) \
= \ 0 \ = \ D_t (J_Y) \ . $$

We also note that while $J_X$ is also conserved under the standard
Euler-Lagrange equations, for $J_Y$ we would get a non-zero time
derivative, given explicitly by $F r^2 \rho^2 z$.

\subsection*{A.3 Gauge transformation}

Finally, we should check that a gauge transformation transforms
our problem with twisted symmetries into one with standard
symmetries.

The form of the Lagrangian $\L$, which we rewrite using the
simplified notation $\b = [(x^2+y^2) z]$ as $$\label{exa:lag2} \L
\ = \ (1/2) \ [\zd + \b \, f(\b ) ] \ \( \xd^2 + \yd^2 \) \ - \ g
( \b ) \ , $$ suggests that it can be written as
$$ \label{exa:lag3} \L \ = \ (1/2) \ [\nabla_t z] \ [ (\nabla_t x
)^2 + (\nabla_t y)^2 ] \ - \ g ( \b ) \ , $$ in terms of covariant
time derivatives, defined as $$ \label{exa:covder} \nabla_t x = d
x/dt , \ \nabla_t y = d y / dt , \ \nabla_t z = (d z / dt)  + \b
f(\b ) \ . $$ In other words, the operator $\nabla_t$ acts on the
vector $(x,y,z)^T$ by
$$ \nabla_t \ = \ \frac{d}{dt} \ + \ \La $$
with $\La$ as given above.

The covariant $t$-derivative of $z$ can be mapped into a standard
$t$ derivative by a change of variables; as we need $$ \nabla_t z
\ = \ z_t + \b f(\b) \ = \ \zeta_t \ , $$ this yields $$ z = \s
\zeta \ \mathrm{with} \ \ \s = \exp \[ - \int \b \, f(\b) \, dt \]
\ . $$ Needless to say, we do not need to change variables for $x$
and $y$; we will introduce new variables $\xi = x$ and $\eta = y$
just for the sake of stressing the different set of variables. In
these variables, we have $$ \label{exa:lag4} \L \ = \ (1/2) \
\dot{\zeta} \ \( {\dot{\xi}}^2 + {\dot{\eta}}^2 \) \ - \ g ( \b )
\ ; \eqno(A.4) $$ note that writing $\b$ in terms of the new
variables is in general a nontrivial task, as the change of
variables depends itself on $\b$.

As the gauge transformation we considered -- and hence the change
of variables needed to set the Lagrangian in the form (A.4) -- do
not act on the $(x,y)$ variables, we can forget about the vector
field $X$ and concentrate on $Y$.

In the new variables the basic differential operators are written
as $ \pa_x = (\pa \xi / \pa x) \pa_\xi + (\pa \eta / \pa x)
\pa_\eta + (\pa \zeta / \pa x) \pa_\zeta$, and so on. This yields
explicitly
$$ \begin{array}{l}
\pa_x \ = \ \pa_\xi \ - \ (z/\s^2) \s_x \, \pa_\zeta \ , \\
\pa_y \ = \ \pa_\eta \ - \ (z/\s^2) \s_y \, \pa_\zeta \ , \\
\pa_z \ = \ [ (1/\s) \, - \, (z/\s^2) \s_z ] \, \pa_\zeta \ .
\end{array} $$
Note however that $\s$ depends on $(x,y,z)$ only through the
function $\b$; thus we have $\s_x = \s_\b \b_x$, and so on. It
follows from this that
$$ X \ = \ - \eta \, \pa_\xi \ + \ \xi \, \pa_\eta \ ; \ \
Y \ = \ \xi \, \pa_\xi \ + \ \eta \, \pa_\eta \ - \ 2 \la (\b) \,
\z \, \pa_\z \ ; $$ here the exact expression of $\la (\b)$ is
inessential, but for the sake of completeness we mention that,
writing $A = \b f(\b)$, it is given by
$$ \la (\b) \ = \ (1/\s) \ \( 1 + \b \int (d A / d \b) dt \) \ .
$$

According to our general discussion, there should be a vector
field $W$, obtained via a gauge transformation from $Y$, which
leaves the Lagrangian $\L$ invariant. This is immediately seen in
the new variables, i.e. with the representation (A.4) for $\L$.
The gauge transformation is simply $ \Ga = [1/\la (\b)] \, M $, so
that $$ W \ = \ \xi \, \pa_\xi \ + \ \eta \, \pa_\eta \ - \ 2 \,
\z \, \pa_\z \ ; $$ needless to say, this is a rescaling in the
new variables, and the invariance of $\L$ under this, or more
precisely its standard prolongation, is immediate.

\end{document}